\documentclass[conference]{IEEEtran}
\IEEEoverridecommandlockouts
\usepackage{cite}
\usepackage{hyperref}
\usepackage{amsmath,amssymb,amsfonts}
\usepackage{algorithmic}
\usepackage{graphicx}
\usepackage{textcomp}
\usepackage{xcolor}

\def\BibTeX{{\rm B\kern-.05em{\sc i\kern-.025em b}\kern-.08em
    T\kern-.1667em\lower.7ex\hbox{E}\kern-.125emX}}
\begin{document}

\title{Mechanism Design and Blockchains
}

\author{
\IEEEauthorblockN{Akaki Mamageishvili}
\IEEEauthorblockA{\textit{Dept. of Management, Technology and Economics} \\
\textit{Swiss Federal Institute of Technology}\\
Zurich, Switzerland \\
amamageishvili@ethz.ch}
\and
\IEEEauthorblockN{Jan Christoph Schlegel}
\IEEEauthorblockA{\textit{Dept. of Economics} \\
\textit{City, University of London}\\
London, UK \\
jansc@alumni.ethz.ch}
}

\maketitle

\begin{abstract}
Game theory is often used as a tool to analyze decentralized systems and their properties, in particular, blockchains. In this note, we take the opposite view. We argue that blockchains can and should be used to implement economic mechanisms because they can help to overcome problems that occur if trust in the mechanism designer cannot be assumed.
Mechanism design deals with the allocation of resources to agents, often by extracting private information from them. Some mechanisms are immune to early information disclosure, while others may heavily depend on it. Some mechanisms have to randomize to achieve fairness and efficiency. 
Both issues, information disclosure, and randomness require trust in the mechanism designer. If there is no trust, mechanisms can be manipulated.  
We survey mechanisms that use randomness or sequential information disclosure and are hard, if not impossible, to audit. Therefore, mechanisms for which centralized implementation is often not a good solution. We consider some of the most frequently used mechanisms in practice and identify circumstances under which manipulation is possible. We propose a decentralized implementation of such mechanisms, that can be, in practical terms, realized by blockchain technology. Moreover, we argue in which environments a decentralized implementation of a mechanism brings a significant advantage.  

\end{abstract}


\section{Introduction}

 Most economic mechanisms implicitly assume trust in the mechanism designer. In this note, we will address two different trust assumptions that economic mechanisms make and argue that blockchains and smart contracts can be used to implement mechanisms even if these assumptions are not met. First, in most mechanisms, agents submit private information to the mechanism designer who uses it as an input to decide on how to allocate goods. However, sometimes, private information must be received simultaneously, and goods have to be allocated instantly. That is, the designer should not provide any information that he has obtained from some agents to the remaining agents to put them into an advantageous position.

Consider, for example, auctions. They are used for allocating or re-allocating goods to agents in numerous applications, including financial markets (bond sales, initial public or security token offerings, opening auctions in stock exchanges), art and antique markets, procurement, advertisement, wholesale, etc. Auctions are run since ancient times, but recently more and more auctions are performed electronically. Therefore, the use of blockchain technology in auctions has become relevant\footnote{See \url{https://medium.com/crypto-economics/an-introduction-to-auction-theory-blockchain-edition-cf09b005b1cc}}.

In auctions, the auctioneer collects bids for the item at sale from bidders, allocates the item, and defines payments. Depending on the auction format (who gets the item and how payments are defined), it might be beneficial for some bidders to learn about other agents' bids.     

Second, some mechanisms use random input bits to decide on the allocation. It is, therefore, important that the designer does not manipulate the source of randomness. 
Consider, for example, the allocation of school seats to students by lottery.
Lotteries are used to allocate school seats and access to charter schools in large school districts like Boston, Chicago, or NYC. 
In these mechanisms, randomization sometimes happens on a school level or on a school district level. In some cases, manual ways are used to generate randomness; in others, computer software is used. 

We identify examples of mechanisms that could benefit from a decentralized implementation. In practical terms, such implementation can be done using already existing blockchain infrastructure using smart contracts. Alternatively, a new blockchain environment can be created to maintain a decentralized database for the input of the mechanism.   

\section{Background}

\subsection{Mechanism Design}

Mechanism design is at the intersection of economics and computer science. It takes an engineering approach to the design of economic mechanisms and incentives. In many situations, agents hold private information that is relevant for determining a welfare-optimizing allocation of resources, or they take unobserved actions. The mechanism designer creates or improves economic institutions that incentivize agents to reveal their private information or to take desirable actions. Participants in the mechanisms often are assumed to be self-interested and strategic. The mechanism designer then has to solve the problem of finding a mechanism that maximizes his objective (e.g.welfare, revenue, fairness) under participation and incentive constraints: agents have to be incentivized to participate in the mechanism and to reveal relevant information, respectively, to take optimal actions.
Alternatively, mechanism design can be seen as a form of reverse game theory, where the mechanism designer designs a game that results in desirable outcomes. Applications are numerous, spanning economics and politics (markets, auctions, computational social choice, monetary policy, insurance, and labor contract design) to computer science (networks, internet routing, sponsored search auctions, security). 

Game theory proposes different solution concepts that aim to predict strategic outcomes. A concept that only makes minimal assumptions on the agents' strategic sophistication and ability to predict other agents' actions is that of a dominant strategy. Applied to the context of mechanism design, in a {\it dominant strategy incentive compatible} (DSIC) mechanism, it is optimal for each player to reveal her private information truthfully, no matter what other agents do or report. However, such mechanisms are rare. Stronger assumptions on the abilities of agents, lead to solutions concepts such as Bayesian incentive compatibility (BSIC) that allow for a wider range of possible mechanisms. However, their analysis assumes equilibrium play and correct beliefs of agents, and hence these mechanisms are sensitive to agents' knowledge and beliefs about other agents. For a systematic treatment of mechanism design, see~\cite{Borgers} and the references therein. Recently, tools developed in the mechanism design literature have been applied in the context of blockchains. See, for example,~\cite{blockRewards}. 

Even if a dominant strategy implementation is possible, we might worry about the incentives of the mechanism designer. The mechanism design literature normally ignores the strategic or malicious motives of the mechanism designer. A recent contribution~\cite{credible_mechanisms} is an exception. We will discuss these issues in more detail later. 

\subsection{Blockchains}

Blockchain was introduced in~\cite{nakamoto}, as a tool to maintain the (crypto-)currency Bitcoin. The invention of the blockchain made it possible to solve the double-spending problem without the need for a trusted or central authority.

By definition, a blockchain is a growing list of records collected in blocks, that are linked by using cryptographic primitives. Each block contains a cryptographic hash of the previous block, a timestamp, and transaction data, which is often represented as a Merkle tree.

An important property of a blockchain is that it makes a modification of the data practically infeasible. To be used as a distributed ledger, a peer-to-peer network maintains a blockchain. Participating nodes collectively adhere to a protocol about node communication and validate new blocks of such communication. The blockchain is open to everyone in the network, using a distributed ledger that efficiently records transactions between two parties, in a verifiable way. 
 
Once recorded, the data in any given block cannot be changed without alteration of all subsequent blocks, which normally requires a consensus of the majority of nodes. 
Even though blockchain records are, in principle, alterable, blockchains can still be considered secure. They are an example of a distributed computing system with a certain degree of Byzantine fault tolerance. The decentralized consensus is the ultimate goal of a blockchain.

Storing the data across a peer-to-peer network eliminates some risks, in particular, risks that come with a central entity holding the data. The decentralized blockchain can use the state of the art message passing and distributed networking technology.
Blockchain security methods include the use of public-key cryptography. A public key signs every transaction. The main property of the public key is that only the owner can sign it. A private key is like a password that gives its owner access to their digital assets and other functionalities that blockchains support. 

Blockchains use different time-stamping schemes to maintain data integrity. The original and still most used such scheme is proof-of-work. 
Alternative consensus methods include proof-of-stake.

Mechanisms can be implemented via {\it smart contracts}. A smart contract, first introduced in~\cite{smart_contracts}, is a self-executing ''programmable" protocol, intended to facilitate, verify, and/or enforce the rules or performance of a contract. Smart contracts allow the execution of credible transactions without trusted third parties. Transactions can be tracked, and they are irreversible. A blockchain-based smart contract is visible to all participants of that blockchain.

\section{Information Disclosure}

\subsection{Examples of Manipulation}

\subsubsection{First-Price Auctions}
First-price sealed-bid auctions are frequently used in practice.
Agents bid on an item, and the highest bidder wins the item and pays her bid. This mechanism can be manipulated by the auctioneer (who acts on behalf of the seller but whose interests are not aligned with the seller) in the following way. The auctioneer can tell the highest bidder the bid of the second-highest bidder, $b_2$. The highest bidder can send the bid $b_2+\varepsilon$, where $\varepsilon>0$ is some positive number. The highest bidder claims the item and pays $b_2+\varepsilon$. The auctioneer and the bidder can share the surplus. The seller loses $b_1-b_2-\varepsilon$ in potential revenue.

First-price auctions are, for example, frequently used for procurement by the government or businesses. If the governmental official conducting the auction is corrupt, as has been documented for first-price sealed-bid procurement auctions in Russia \cite{RussianCorruption,MoreRussianCorruption}, he can extract rents' by strategically revealing information, as described above.

\subsubsection{Second-Price Auctions}
Second-price auctions are also frequently used in practice.
Agents bid on an item, and the highest bidder wins the item and pays the second-highest bid.
 This mechanism is the dominant strategy incentive compatible. However, there is room for manipulation if the information is not disclosed simultaneously. The auctioneer (whose interests can now be perfectly aligned with the seller) can, after seeing the bid from the highest bidder, approach the second-highest bidder and ask to increase her bid right below the highest bid. The second highest bidder does not lose anything, while the auctioneer can share the gains. Mechanisms that can not be manipulated in this way by the seller are called credible in~\cite{credible_mechanisms}. The paper shows that the first-price auction is the unique credible static mechanism. In particular, the second-price auction is not credible, as we have outlined above. Although the first-price auction is credible in this sense, we have seen in the previous section, that it also suffers from the possibility of malicious information disclosure, depending on the context.

\subsubsection{Generalized Second-Price Auctions}

Generalized second-price auctions are used in practice as well, in particular for sponsored search. There are $n$ agents and $k<n$ slots, for each keyword, to be allocated. Each slot $i$ has a click-through rate, $\alpha_i$.  Without loss of generality assume that $\alpha_i>\alpha_{i+1}$ for any $i\leq k-1$. Denote the bid of agent $i$ by $b_i$ and suppose that bids are sorted. 

We distinguish between two cases. In the first, the auctioneer's interests are perfectly aligned with the seller of the slots. Then, he can communicate information to the agent with the $k+1$-highest bid, which does not get a slot. If the agent increases her bid to $b_k-\varepsilon$, she does not get the $k$-th slot, but the auctioneer can share the surplus gained from the $k$-highest bidder, similarly to the case of the second-price auction. 

In the second case, the auctioneer's interests are not aligned with the seller of the slots. Then, he can disclose information to a high bidder about more attractive lower slots. In this case, the high bidder can deviate by bidding a lower number. Consider the following examples. There are $2$ slots and $3$ bidders. The click-through rates are $\alpha_1=1.0$ and $\alpha_2=0.8$. Bids are $b_1=10$, $b_2=9$ and $b_3=1$. 
If the highest bidder learns about the lowest bid, she is incentivized to bid between $1$ and $9$. In case of truthful revelation, her utility is $10\cdot 1.0-9\cdot 1.0 = 1.0$. In case of a deviation, her utility becomes $10\cdot 0.8-1\cdot 0.8=7.2$, an increase of $6.2$. 

\subsubsection{Boston School Choice Mechanism}
The Boston or Immediate Acceptance Mechanism is a popular method to allocate seats in public schools or universities to students. Students are prioritized according to some criteria, e.g., based on exam score results, or randomly through a lottery (more on that later). Students submit a ranking of their most preferred schools to the mechanism which takes these and the priority information into account to calculate an assignment: First, the mechanism tries to assign each student to her first choice. In the case of over-demand, i.e., if more students have a particular school as the first choice than seats are available, then the seats are rationed according to the priorities with the highest priority students being admitted. Next, the mechanism tries to assign each student who has not been assigned to her first choice previously to her second choice. In case of over-demand, i.e., if more students have a particular school as the second choice than seats are available (subtracting the seats already allocated in the first round as well), then the seats are rationed according to the priorities with the highest priority students being admitted, and so on. This procedure continues until all students have been assigned, or no more school seats remain unfilled.

In general, this procedure is not dominant strategy incentive compatible: Suppose there are two universities, Oxford and Cambridge, with one seat to fill each, and three applicants, Alice, Bob, and Carol. Both Oxford and Cambridge rank Alice before Bob and Bob before Carol. Suppose both Alice and Bob prefer Oxford to Cambridge and Carol prefers Cambridge to Oxford. In the first round of the mechanism, we would try to match both Alice and Bob to Oxford. However, Oxford has only one seat available, which is allotted to the better student, Alice. Carol gets a seat at Cambridge in the first round, because she is the only applicant who ranks Cambridge first. In the second round of the mechanism, we will try to assign Bob to his second choice Cambridge. However, there is no more seat available at Cambridge, as we have previously assigned the only seat to Carol.

Now suppose, the university admissions committee sells the information to Bob that Alice ranks Oxford first, and Carol ranks Cambridge first. Then Bob can gain from this information by strategically reporting Cambridge as his first choice and will obtain a seat at that university.

For more criticism of the mechanism, see~\cite{boston_mechanism}.

\subsection{Mechanisms with Privacy}
\label{privacy}
Recently, economists have highlighted the importance of privacy-preservation in mechanisms,~\cite{transparency},~\cite{menu_mechanisms}.  Participants send their private preferences/types to the mechanism designer, a trusted third party, which takes this information as an input to the predefined algorithm. Such an approach can create mistrust in the mechanism designer, and therefore, in the final allocation calculated by the algorithm. Ideally we want to design mechanisms that preserve privacy while, at the same time, maintaining the possibility to audit the mechanism if a participants has doubts that the mechanism designer correctly applies the allocation algorithm.

\section{Randomness}
Randomness is an important ingredient in designing mechanisms to allocate public and private goods. Its primary function is to achieve fairness: Everything else being equal, if a resource has to be assigned to a person rather than another person and money cannot be used to compensate, then randomly allocating that resource is fair from an ex-ante point of view.

In this note, we want to make the point that the possibility of implementing random mechanisms with trustworthy randomness is actually one of the major advantages of blockchains and smart contracts and can not be easily substituted by a centralized mechanism.

Think about the following situations: Public duties or offices have to be assigned. Frequently this is done randomly. A lottery was used to draft soldiers for the Vietnam war (There seems to be evidence that the outcome of the 1970 lottery was not entirely random). Jury duties are assigned randomly. In ancient Athens and Renaissance Florence, public offices were allocated based on a lottery among eligible citizens. A lottery assigns green cards in the United States and working visas in many countries. Recently there have been proposals for including randomness in democratic decision-making processes, e.g., random sample voting,~\cite{RSV}, assessment voting~\cite{AV}, some of which have been suggested to be used in blockchain governance. Finally, proof-of-stake consensus protocols themselves use randomness to select a subgroup of participants to perform certain tasks.

Using randomness requires trust. If a central authority implements randomness, one has to make sure that bribery does not occur, or that that authority does not manipulate the process in its favor. 
On the other hand, verifiable randomness from a natural source is hard to obtain. Suppose, for example, that we use the weather as a source of randomness. First of all, we have to agree on a weather-related event that determines the outcome. Our choice of such an event depends on our prior information about, e.g., the likelihood of rainfall tomorrow. If we hold different information, it is generally hard to agree on a particular event. 
~\cite{aumann} famously proved that we could not agree to disagree even if we hold different information. However, that presupposes that we are non-strategic, i.e., we don’t have a vested interest and are only concerned with finding the truth.
Suppose we want to implement $x$ if it is raining tomorrow and $y$ if it is not. What are the times and places where we measure the rainfall? 
What is a sufficient amount of precipitation to declare that it rains? Etc. Suppose we have the same prior information and have agreed on the definition of rain and a place where rainfall is measured. Even then, to determine if it is raining tomorrow, in case of a disagreement, we need a third party (an oracle) to measure the rainfall. However, then there might be incentives to bribe that third party. Similar arguments hold for any predefined source of randomness.

This problem is not only recent or theoretical. Bribing frequently occurred in the lottery in renaissance Florence, to the extent that offices were sold rather than randomly allocated. In the design of proof-of-stake consensus protocols, it is a mostly unsolved problem to find a reasonable protocol for determining a token whose holder is eligible to add a block.

The problem of trust is particularly pertinent to lotteries. Deterministic mechanisms are easy to audit. Even if a central authority implements the mechanism, we can reproduce all its steps and make sure that the reported outcome is correct. Thus in many circumstances, it is not necessary to proceed in a decentralized way. In lotteries, however, auditing is not easily possible. Still, we can achieve trust in a distributed way on a smart contract, as we describe next.

\section{Implementation}

We briefly discuss one simple, but robust, way to design a decentralized random number generator as a smart contract, in a blockchain environment, e.g., such as Ethereum,~\cite{wood2014ethereum}. We generate a random number in two phases. 
In the first phase, each player $i$ generates her (pseudo) random number $c(i)$ and then commits to $c(i)$ by sending the hash $h(c(i))$ to the contract. $h$ should be a cryptographic hash function. 
Cryptographic hash functions should satisfy several properties, of which we detail the most relevant ones to this approach. 
First, the same input sequence should always result in the same output hash. Second, it should be infeasible to generate a message from its hash (i.e., the hash should not reveal any information about the input to the hash function).
Third, it should be infeasible to find so-called hash collisions. A hash collision occurs when two different input sequences are found, which result in the same hash value. The security of cryptographic hash functions is typically defined by how well they resist collision attacks, depending on which input sequence the adversary can choose from (any input corresponds to preimage resistance, a fixed input corresponds to second preimage resistance).
Additional desirable properties are the following: it is quick to compute the hash value for any given message, and a small change to a message should change the hash value extensively. That is, the new hash value should look uncorrelated with the old hash value. For the references, see~\cite{goldwasser1996lecture,rivest1992md5,eastlake2001us,standard2002fips, SHA3}.  

The first phase ends after a predetermined amount of time $T$. In the second phase, each player announces the actual random number $c(i)$ she has generated, by sending it to the contract. The contract can check whether the submitted number has the right hash sent previously: If the hash is not correct, the contract punishes the player by excluding her from the game. In the context of allocating goods, the punishment would preclude her from receiving any good. Assuming that the hash function h is one-way, no player can change her number after having committed to it. The second phase ends after a predetermined amount of time $T'$, and then the contract computes the sum of all available numbers to use it as the final random number. It is easy to note that with just one of the players playing honestly, the final output is random, as desired.

The approach described above turns out to be robust. A player cannot gain from delaying to uncover her random number in the second phase. If she waits for other players to reveal their numbers and does not reveal her number until $T’$, she does not receive a good. Similarly, corrupting any strict subset of players does not make a difference. The algorithm can only be compromised through the collusion of all participating players. In the latter case, however, we assume that there is no solution needed in the first place. 

Moreover, in contrast to a centralized procedure, it only requires that one participant does not collude without needing to know ex-ante who that non-colluding participant is. The other source of manipulation is on the side of miners. A miner can manipulate the process by not processing some players’ messages in the second phase. For avoiding this type of manipulation, we should make the second phase long enough to guarantee that all players will add their initial numbers to the blockchain. Similar protocols have been implemented in a blockchain environment, e.g., in Randao\footnote{See the whitepaper~\cite{randao}.}. For a cryptographic analysis of similar protocols, see~\cite{implementation}. A recent trend in literature on {\it Verifiable Delay Functions} also deals with verifiable lotteries. However, additional assumptions on the computing hardware possibilities are required, see~\cite{vdf_original,vdf_simple,vdf_efficient}. 

\begin{figure*}[t]
\centering
\includegraphics[width=13.5cm]{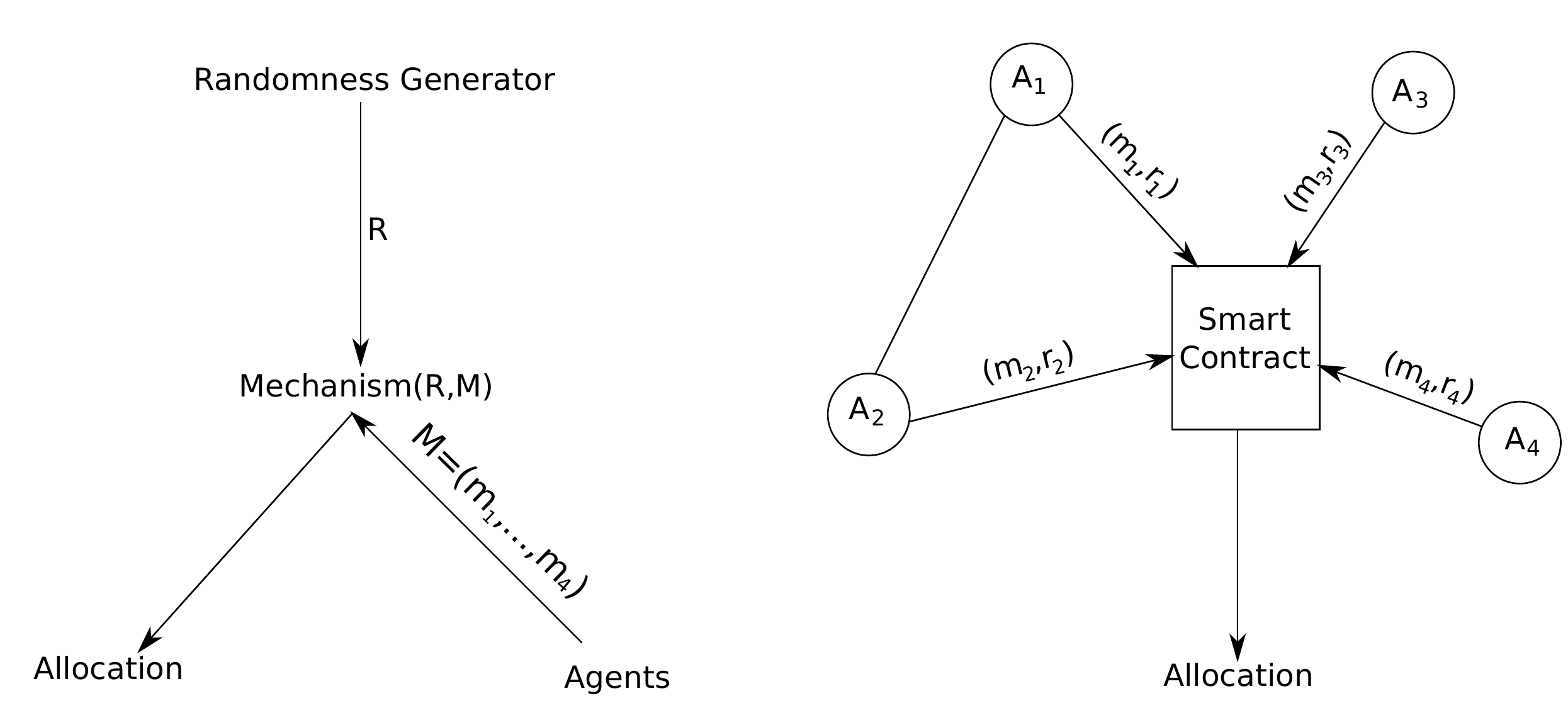}
\caption{In a centralized implementation (left figure) of a mechanism, the mechanism designer receives a vector of messages $M$ from the agents and a string of random bits from an external source to generate an allocation. In a decentralized implementation (right figure), agents form a peer-to-peer network, and each agent $i$ sends a message $m_i$ and a random bit $r_i$ to a smart contract, which calculates an allocation. }
\end{figure*}

Similarly, players of the mechanism can send hashed private information to a mechanism, $h(p_i)$ and disclose $p_i$, after the first round. In the context of auctions, $p_i$ is the bid of an agent. In case of school choice mechanisms, $p_i$ is the ranking of all schools for each student, that is the vector $(s_1,s_2,\cdots,s_n)$, where $s_i$ is the $i-$th most preferred school. This way, the mechanism gets all the data from the players in a simultaneous way, and the disclosure of some information unless the player wants it, does not take place. 

Every agent in this decentralized system can have a copy of the blockchain. Extensive database replication and computational trust maintain data quality. No centralized ''official" copy has to exist, and no user has to be ''trusted" more than any other.

For the deferred acceptance mechanisms described in Section~\ref{privacy}, a general solution could be multiparty computations as developed in~\cite{gen_multiparty}. The paper offers a way to calculate any function $f(x_1,\cdots, x_n)$ given private information $x_i$ of the $i$-th participant, without disclosing any information besides the obvious one inferred from the final allocation. The only two assumptions are the existence of {\it one way functions} and that majority of the processing nodes are honest. However, general computational and communication complexity can be high depending on the particular form of the function $f$. For the particular case of matching problems, which are typically solved using the Deferred Acceptance algorithm,~\cite{first_efficient_mtch} has developed a privacy preserving and practically implementable protocol. Later, the protocol was corrected and improved in~\cite{improved_efficiency}, while~\cite{impl_efficient} implemented the protocol as a mobile application.

We suggest that, if the mechanism is simple enough, it should be implemented as a blockchain itself. For example, the Boston school choice mechanism and auctions discussed above fall into a class of simple mechanisms. In the blockchain implementation, transactions, in this context, messages containing random numbers or private information, can be broadcast to the network using the blockchain software. Mining nodes can validate transactions, add them to the block they are building, and then broadcast the completed block to other nodes.

Private or permissioned blockchains can be of particular interest in our context. 
Permissioned blockchains have an extra layer for access control. At this layer, the network decides to give access to a new node. The agents in the game can be the nodes of the network. Starting node can be a central authority, that invites/adds all agents to a network. After the very beginning, the role of a central authority is minimal.

On the other hand, if the mechanism is more involved, for example, a course scheduling mechanism~\cite{course_matching}, then it can be implemented as the smart contract. However, some measures of precaution have to be taken. For instance, Ethereum based smart contracts need {\it Gas}, a fraction of an Ethereum token, for the execution. Each operation of a smart contract has its cost in Gas. Therefore, very complicated mechanisms might turn out to be too expensive to implement as a smart contract.~\cite{double_deposit} presents one example of an economic mechanism implementation as a smart contract.

\section{Conclusion}

We have argued that implementing economic mechanisms that allocate goods to agents is a perfect use-case of blockchains. Mechanisms that use randomness or that are sensitive to early information disclosure can be implemented in existing blockchains environments, in the form of smart contracts, or by creating a new blockchain environment, where the nodes in the network are the participants of the mechanism.
As we argued, exact implementation details are mechanism dependent, and we leave them for future research. 

\bibliographystyle{plain}
\bibliography{sample}

\end{document}